\documentclass[aps,prb,showpacs,twocolumn,floatfix]{revtex4}
\usepackage{graphicx}

\begin{document}
\title{A $dc$ voltage step-up transformer based on a bi-layer $\nu =1$ quantum Hall
system}
\author{B. I. Halperin$^1$, Ady Stern$^2$, S. M. Girvin$^3$}
\affiliation{$^1$Department of Physics, Harvard University, Cambridge, MA 02138\\
$^2$Department of Condensed Matter Physics, Weizmann Institute of Science,
Rehovot 76100, Israel\\
$^3$ Sloane Physics Laboratory, Yale University, P.O.~Box 208120, New Haven,
CT 06520-8120}

\begin{abstract}
A bilayer electron system in a strong magnetic field at low temperatures,
with total Landau level filling factor $\nu =1$, can enter a strongly
coupled phase, known as the (111) phase or the quantum Hall
pseudospin-ferromagnet. In this phase there is a large quantized Hall drag
resistivity between the layers. We consider here structures where regions of
(111) phase are separated by regions in which one of the layers is depleted
by means of a gate, and various of the regions are connected together by
wired contacts. We note that with suitable designs, one can create a {\it dc}
step-up transformer where the output voltage is larger than the input, and
we show how to analyze the current flows and voltages in such devices.
\end{abstract}
\maketitle

\section{Introduction}
Following earlier theoretical predictions
\cite{duan,smgdrag,kyangdrag,kimdrag,smgNobel}, recent
experiments have revealed \cite{kellogg-drag} a unique
behavior of coupled electronic transport in a bilayer
electronic system in the quantum Hall regime, when the two
layers have a total Landau level filling factor $\nu =1$
(each layer separately being at $\nu \approx 1/2$). If the
separation between the layers is sufficiently small, relative
to the distance between electrons in a layer, the system can
enter a strongly coupled state at low temperatures known as
the $(111)$ or quantum Hall pseudospin-ferromagnet phase
\cite{smgdrag,DasSarmaPinczuk,SMGleshouches,kyang}. It was
predicted for this phase, that if there is no tunneling
between the layers, and a current $I$ is driven in one of the
layers (the ``active layer''), with no net current flowing in
the other (``passive'') layer, then the voltage drop should
be identical in the two layers. In the limit of zero
temperature, this voltage drop should be purely perpendicular
to the current, and equal to $I\frac{h}{e^{2}}$ in each
layer. Experiments have confirmed this quantization of the
Hall drag resistance with an accuracy of order $10^{-3}.$

The properties of the $(111)$ phase reflect a novel form of
interlayer phase coherence, which may be understood as a kind
of superfludity in the difference of the electric currents in
the two layers, and which shorts out any differences in the
electric fields within the two layers
\cite{duan,smgdrag,kyangdrag,kimdrag,smgNobel}. The coherent
state has a broken symmetry which leads to a Goldstone
collective mode \cite{fertig} and to a giant zero-bias
anomaly in the interlayer tunneling spectrum
\cite{wenzee,fogler,balents,stern}, which have both been
observed experimentally in Eisenstein's group
\cite{JPEgoldstone,JPEtunnelanomaly}.

In this work we use this equality of the Hall voltage between
the two layers to show that a properly constructed bi-layer
system, incorporating regions of the $(111)$ phase separated
by regions where one of the layers is depleted, may serve as
a $dc$ voltage step-up transformer.\cite {halperinxfrmr} More
generally, we show how to analyze the current flows and
voltages in devices made up of alternating regions containing
the $(111)$ phase and regions where one or the other layer is
depleted by a top or bottom gate. We assume throughout that
there is no tunneling between the two layers, either because
the barrier is too high, or because tunneling has been
supressed by application of a parallel magnetic field.

\begin{figure}
\includegraphics[width=3.375in,angle=0,clip=]{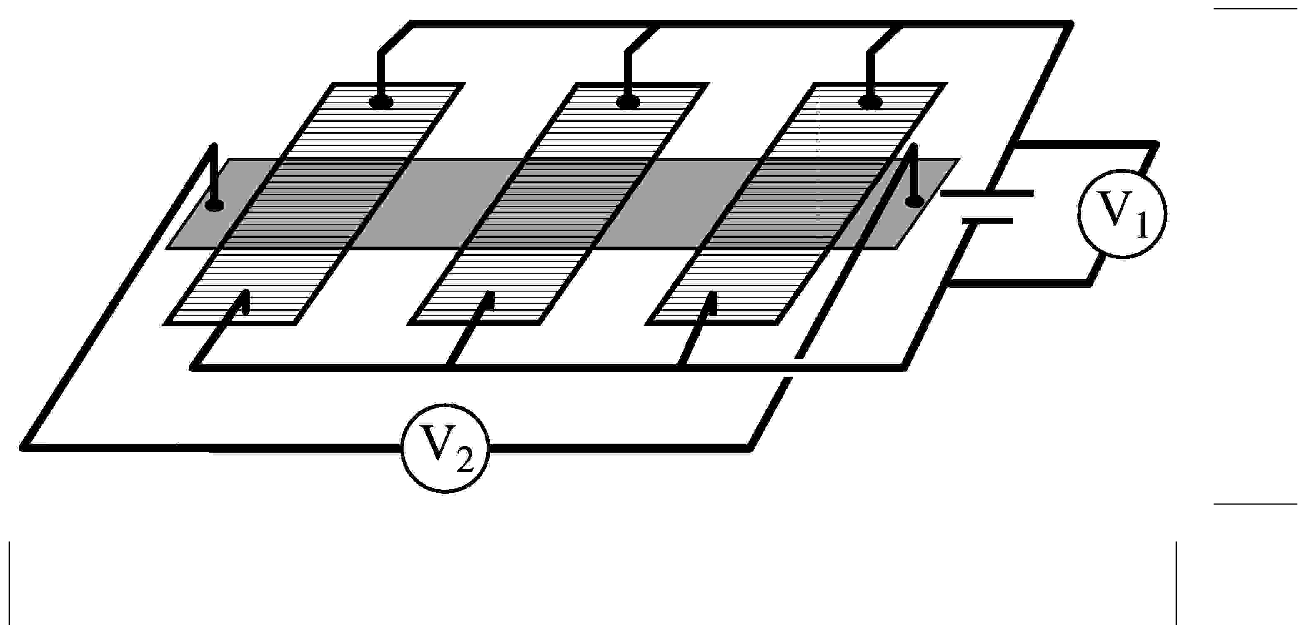}
\caption{\label{1}One version of the transformer, with $N=3$ stages. Horizontal
stripes indicate regions where the upper layer is occupied; shading
indicates regions where the lower layer is occupied. Areas with both stripes
and shading have both layers occupied, with the system in the strongly
coupled (111) phase at total Landau-level-filling $\nu =1$. The upper layer,
divided into strips connected in parallel, is used as the primary. If a
current $I_{1}$ flows in each strip, then a voltage $V_{2}=NI_{1}h/e^{2}$ is
induced in the secondary layer, provided no current is drawn.}
\end{figure}
In 1965 Ivar Giaver realized a {\it 1:1} dc transformer using flux flow
resistance in magnetically coupled superconducting layers. \cite{giaver} In
QHE bilayers the layer coupling is of electrostatic origin but the
transformer action can be viewed within the composite boson picture as
arising from the flow of Chern-Simons flux attached to the particles. In the
case of a superconductor, flux flow is mostly perpendicular to the current
flow while in the present case the Chern-Simons flux is attached to the
particles themselves and flows with them, inducing a voltage drop in the
secondary perpendicular to the current flow in the primary. \cite{smgNobel}

The proof of concept is most simply seen in the geometry of Fig. [1], where
the upper active layer is used as the primary circuit, with $N$ primary
strips in parallel. A time-independent voltage $V_{1}$ will lead to a
time-independent current $I_{1}$ in each of the strips, and this current is
independent of $N$. The voltage across the passive layer, which acts as a
secondary circuit, will be $V_{2}=NI_{1}h/e^{2}$ when no current is drawn.
Clearly, we will have $V_{2}$ larger than $V_{1}$ if $N$ is sufficiently
large. As we show below, the output impedance of the transformer is non-zero
so that if a non-zero current $I_{2}$ is drawn from the secondary, the
secondary voltage $V_{2}$ will decrease, and the primary voltage $V_{1}$
will increase, for fixed $I_{1}$. Nevertheless, if the current drawn is
sufficiently small, the secondary voltage will remain larger than the
primary.

Another possible geometry, having the strips connected in
series rather than parallel, is illustrated in Fig. [2]. A
time-independent current $I_{1}$ flows in the primary layer,
denoted layer $1$. This current is driven by a battery with
voltage $V_{1}$. The voltage $V_{2}$ measured in the
secondary
circuit, when no current is drawn, is equal to $NI_{1}h/e^{2}$ for an $N$%
-stage device. (Again, the voltage will be reduced when finite current is
drawn from the secondary.) In this case, the voltage $V_{1}$ in the primary
circuit is also proportional to $N.$ However, we show below that with proper
design, the voltage drop $V_{1}$ will be smaller than $V_{2}$, so that
voltage gain is achieved. By connecting together several devices, with the
secondaries in series and the primaries in parallel, one can obtain an
arbitrarily large multiplication factor for the voltage.

\begin{figure}
\includegraphics[width=3.375in,angle=0,clip=]{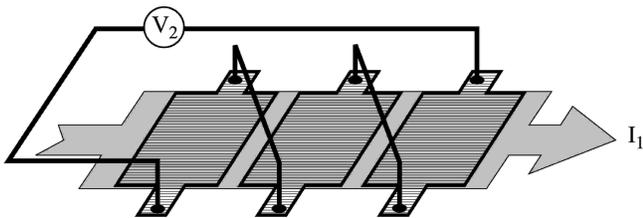}
\caption{\label{2}Alternate version of the transformer. Here the lower layer is used
as the primary, while the upper layer, divided into strips connected in
series, is used as the secondary.}
\end{figure}

The structures we discuss are inherently non-uniform, since one of the
layers is depleted in parts of the sample. As a consequence, the analysis of
current flows and voltage drops is non-trivial. Below, we shall first carry
out such an analysis for the case where no current is drawn from the
secondary, and then consider the case where $I_{2}\neq 0$. We will mostly
analyze the device shown in Fig. [2], and discuss the device in Fig. [1]
towards the end of the paper. We confine ourselves to the situation where
the dimensions of the transformer are large compared to any relevant
microscopic length, including the mean-free-path of any charge-carriers. We
can then use macroscopic conductivity laws and Kirchoff's equations to
determine the current flows and voltage drops in each layer. It is important
to distinguish the classical and quantum aspects of our calculation: the
striking transport properties of the $(111)$ phase, particularly the
existence of a Hall voltage in a layer where no current is flowing, are a
consequence of the quantum Hall effect. In contrast, the non-uniform current
distribution we find in some of the regimes we consider, and particularly
the confinement of dissipation to ``hot spots'', is a consequence of the
classical Kirchoff's laws for non-uniform systems in a magnetic field.

\section{Resistances and current flows}

Let us define $R_{N}$ as the ratio $V_{1}/I_{1}$ for an
$N$-stage device of the type shown in Fig.~[2], when no
current is drawn from the secondary. In
the limit where $N$ is large, we can ignore end effects, and write $%
R_{N}=NR^{\ast }$, where $R^{\ast }$ is a constant, assuming that all
intermediate stages are identical to each other. The value of $R^{\ast }$
can be calculated by considering an infinite periodic system, with the unit
cell shown in Fig.~[3]. The primary layer occupies a region of width $w$
running parallel to the $x$-axis, which we label $0<y<w$. The secondary
layer is depleted in a short region, of length $L_{d}$, which we take here
to be the region $-L_{d}/2<x<L_{d}/2$. In the remainder of the unit cell, of
total length $L_{c}$, both layers are occupied, and the system is in the
(111) phase. We have assumed that the voltage tabs attached to the secondary
layer shown in Fig.~[2] are narrow, so they do not perturb the current flow
when no net current is drawn from the voltage contacts. Thus there is no
current flow in either layer across the boundaries at $y=0$ and $y=w$.

We assume that in the parts of the sample where one of the layers is
depleted, the other layer (whose filling factor is approximately $1/2$) is
in a compressible state, characterized by a resistivity tensor $\rho ^{1}$
with
\begin{eqnarray}
\rho _{yx}^{1} &=&-\rho _{xy}^{1}\equiv \nu _{1}^{-1}\approx 2 \\
\rho _{xx}^{1} &=&\rho _{yy}^{1}\equiv \epsilon \ll 1
\end{eqnarray}
We have chosen the magnetic field direction along the positive $z$-axis, so
that $\rho _{yx}^{1}$ is positive, and we use units where $h/e^{2}=1$ for
intermediate steps of the calculation. In the coherent $(111)$ region, where
both layers are occupied, there is no resistivity to a flow of an
anti-symmetric current. For a symmetric current the Hall resistivity is
quantized $\rho _{yx}^{{\rm coh}}=-\rho _{xy}^{{\rm coh}}=1$. The diagonal
resistivity vanishes rapidly at low temperatures and we take it here to be a
negligibly small positive quantity. We carry out our analysis of the current
flow patterns assuming that the diagonal resistivities in each of the two
phases do not fluctuate with position. However, our results for the
operation of the device as a $dc$ voltage step-up transformer are
independent of that assumption.

Let ${\bf j}^{\alpha }({\bf r})$ and $\phi ^{\alpha }({\bf r})$ be the
current density and the potential in layer $\alpha ,$ and ${\bf E}^{\alpha
}=-{\bf \nabla }\phi ^{\alpha }$ be the electric field in layer $\alpha .$
In the geometry of Fig.~[2] no current flows in the secondary layer, i.e., $%
{\bf j}^{2}=0$. This is obviously true in the regions where this layer is
depleted, but in fact holds also in the coherent $(111)$ regions. In these
regions, the superfluidity shorts out any anti-symmetric electric fields, so
${\bf E}^{1}={\bf E}^{2}$, and the potentials in the two layers differ only
by a constant. The anti-symmetric current in the coherent region, ${\bf j}%
^{1}-{\bf j}^{2}$ , is a supercurrent, so its curl and divergence both
vanish \cite{curlfootnote}. In fact, the same holds for the total
(symmetric) current: the divergence ${\bf \nabla \cdot ({\bf j}^{1}+{\bf j}%
^{2})}=0$ due to current conservation, and the curl vanishes since ${\bf %
\nabla \times E}^{1}=0$ and the resistivities in the $(111)$ phase are
independent of position. Thus the values of both ${\bf j}^{{\rm 1}},{\bf j}^{%
{\rm 2}}$ in the coherent regions are determined by the boundary conditions:
their normal component must vanish at the top and bottom edges, $y=0$ and $%
y=w$, while the normal component of ${\bf j}^{{\rm 2}}$ must vanish also at
the boundaries between the coherent and depleted regions. Thus, the current $%
{\bf j}^{{\rm 2}}$ vanishes everywhere, and no current flows in the
secondary layer.
\begin{figure}
\includegraphics[width=3.375in,angle=0,clip=]{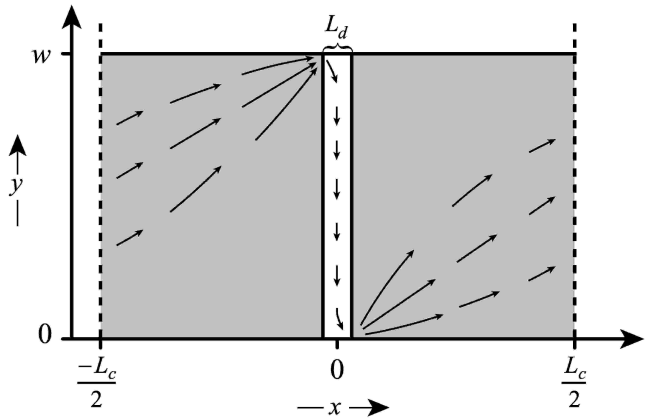}
\caption{\label{3}Schematic unit cell of device shown in Fig 2 lies between vertical
dotted lines at $x = \pm L_c/ 2 $. Shaded region contains the (111) phase,
where both layers are occupied; unshaded region between lines at $x= \pm L_d
/ 2$ has the upper layer depleted. Arrows suggest the flow pattern of an
``extra'' inhomogeneous current, resulting from the $y$-direction current in
the depleted region. To this must be added a uniform current in the $x$%
-direction, so that the total current is $I_1$. The inhomogeneous current is
small compared to the uniform current in regime (i), $L_d/w \ll \epsilon$,
where $\epsilon h/e^2$ is the longitudinal resistivity in the depleted
region.}
\end{figure}

The current distribution ${\bf j}^{{\rm 1}}$ in the active layer may be
analyzed by solving Kirchoff's equations. It is useful to define a ``reduced
potential'' $V({\bf r})$ by
\begin{equation}
{\bf \nabla }\phi ^{1}={\bf \nabla }V-\hat{z}\times {\bf j}^{1}.
\end{equation}
$V({\bf r})$ is the potential corresponding to a current density ${\bf j}%
^{1}({\bf r})$ in a system where the coherent regions are superfluids for
{\it both} symmetric and anti-symmetric currents, and where the depleted
regions have a Hall resistivity of $\nu _{1}^{-1}-1$ rather than $\nu
_{1}^{-1}.$ Within the depleted region
\begin{equation}
\nabla ^{2}V=0,  \label{lapv}
\end{equation}
and four boundary conditions should be imposed. The first two are
\begin{equation}
\epsilon \partial _{x}V+(1-\nu_1 ^{-1})\partial _{y}V=0\, , \text{ \ \ \ \
at \ \ }y=0\text{ and }y=w \, ,  \label{bcy}
\end{equation}
which assure that the current at the edges is parallel to the edges. The
other two,
\begin{eqnarray}
V(x,y) &=&V_0 \, , \text{ \ \ \ \ at \ }x=-\frac{L_{d}}{2} \, ,
\label{bcleft2} \\
V(x,y) &=&0 \, , \text{\ \ \ at \ }x=\frac{L_{d}}{2} \, ,  \label{bcright2}
\end{eqnarray}
result from the vanishing longitudinal resistivity at the $(111)$ region.
The value of $V_{0}$ is proportional to $I_{1},$ and should be chosen such
that
\begin{equation}
\int_{0}^{w}dy\, j_{x}^{1 }= I_{1} \, ,\text{ \ }  \label{bccurrent}
\end{equation}
where the current density is given by
\begin{equation}
{\bf j}^{1}({\bf r})=\frac{\nu _{1}^{-1}-1}{\epsilon ^{2}+\left( 1-\nu
_{1}^{-1}\right) ^{2}}\hat{z}\times {\bf \nabla} V - \frac{\epsilon }{%
\epsilon ^{2}+\left( 1-\nu _{1}^{-1}\right) ^{2}}{\bf \nabla} V \, .
\label{jV}
\end{equation}
The integral in (\ref{bccurrent}) may be taken at any convenient value of $x$%
. Having defined the equation and boundary conditions for $V(r),$ we notice
that $V(r)$ is the potential generated by a capacitor subject to a voltage $%
V_{0}$ and to unusual boundary conditions at the edges.

As we now show, there are three different regimes for this problem,
according to the ratio of the two dimensionless parameters in our definition
of the problem, the aspect ratio $\frac{L_{d}}{w}$, and the longitudinal
resistivity $\epsilon .$ The regimes are: $(i)$ $L_{d}/w\ll \epsilon ;$ $%
(ii) $ $\epsilon \ll L_{d}/w\ll 1/\epsilon $ and $(iii)$ $1/\epsilon \ll
L_{d}/w.$

\subsection{Regime (i): $L_d/w \ll \protect\epsilon$}

We start with the first regime. When $L_{d}/w\rightarrow 0$ the system is
very wide and the edges may be neglected. Far from the edges $j_{x}^{{\rm 1}}
$, $E_{y}^{1}$ are independent of position. They are then equal to $I_{1}/w,$
both in the $(111)$ region and in the depleted region, while $\partial _{y}V$
vanishes. In the $(111)$ region, the electric field is purely perpendicular
to the current, but this is not true in the depleted region, where the
longitudinal resistivity is non-zero. Substituting these values of $j_{x}^{%
{\rm 1}}$, $E_{y}^{1}$ in the relation ${\bf E}=\rho ^{1}{\bf j}$ we obtain,
\begin{eqnarray}
j_{y}^{{\rm 1}} &=&\nu _{1}\left[ \epsilon \frac{I_{1}}{w}-E_{x}^{1}\right] =%
\frac{I_{1}}{w}\frac{1-\nu _{1}^{-1}}{\epsilon },  \label{currenti} \\
E_{x}^{1} &=&\frac{I_{1}}{w}\left\{ \epsilon -\frac{1-\nu _{1}^{-1}}{\nu
_{1}\epsilon }\right\} .  \label{fieldi}
\end{eqnarray}
and $V_{0}=E_{x}^{1}d.$ These results apply throughout the
depleted region, except for small regions, with dimensions of
order $L_{d}$, close to the top and bottom edges, at $y=0$
and $y=w$. (See Subsection \ref{xoversec}, below.) The
integrated current $I_{y}^{1}=L_{d}j_{y}^{1}$ leaves the
depleted region near $y=0$, and spreads out into the $(111)$
region, where it eventually
flows up towards the upper edge and back into the depleted region near $y=w$%
. This extra current flow is shown by the arrows in Fig.~[3].
The integrated current flow across the midline $y=w/2$ in the
(111) region, must be exactly equal and opposite to the
integrated vertical current $I_{y}^{1}$ in the depleted
region, as there can be no net current flow in the
$y$-direction. Since a current density $j_{y} $ at a point in
the $(111)$ region must be driven by a Hall electric field in
the $x$-direction, we see that $\int
E_{x}^{1}dx$ along the midline of a $(111)$ region, say from the point $%
x=L_{d}/2$ to the point $x=L_{c}-L_{d}/2$, must be equal to $I_{y}^{1}$.
Adding in the contribution from the field $E_{x}$ in the depleted region, we
see that the total voltage drop along the midline of a unit cell, say from
the point $x=-L_{d}/2$ to $x=L_{c}-L_{d}/2$ is equal to $I_{1}R^{\ast }$,
with 
\begin{equation}
R^{\ast }=\frac{L_{d}}{w\epsilon }\left[ (1-\nu _{1}^{-1})^{2}+\epsilon ^{2}%
\right].  \label{Rstar}
\end{equation}
We see that $R^{\ast }$ can be made arbitrarily small by using a sample with
a large width $w$, and by making the length $L_{d}$ of the depleted region
as small as possible.

Since there is no current flow across the edges at $y=0$ and
$y=w$, we see that $E_{x}^{1}$ and $E_{x}^{2}$ both vanish
along these edges in the $(111)$ regions. The potentials
$\phi ^{\alpha }$ are therefore constants along each of these
edges, in a given $(111)$ region, and the potential
difference between $y=0$ and $y=w$ in a given layer is just
the Hall voltage, $I_{1}$. The difference in potential
between layer 1 and layer 2 is an arbitrary constant that has
no effect on the current flows or voltage drops along the
layers. Thus, if the edges of the secondary layer are
connected together as shown in Fig.~[2], and no current is
drawn from the secondary circuit, we obtain
$V_{2}=NI_{1}h/e^{2}$ as claimed in the introduction.

\subsection{Regime (ii): $\protect\epsilon \ll L_d/w \ll \protect\epsilon^%
{-1}$}

Naively one might expect the analysis above to hold as long as $L_{d}\ll w,$
but in fact this is not the case. As often happens at the interface of
materials with different Hall resistivities in the regime of strong magnetic
fields, the current distribution can become very inhomogeneous, when the
Hall angle is large, and this can have a major effect on the voltage drop
(see, e.g., \cite{ruzin}, and references therein). However, the analysis
again becomes simple in regime (ii), where $\epsilon $ is small compared to
both $L_{d}/w$ and $w/L_{d}$. We can understand this regime by taking the
limit $\epsilon \rightarrow 0$ with $L_{d}$ and $w$ fixed. In this case the
boundary conditions (\ref{bcy}) - (\ref{bcright2}) imply that the reduced
potential $V(x,y)$ is a constant along each boundary of the depleted region,
with the exception of two ``hot spots'', at corners where two boundaries
meet. At these corners, there is a discontinuity in $V$, or more accurately,
a very rapid change of the potential, on a length scale of order $\epsilon
L_{d}$. In the limit $\epsilon \rightarrow 0$ the resulting divergence of
the electric field leads to a finite total current crossing the boundary
between the (111) region and the depleted region in the corner, with a
finite amount of dissipation. Everywhere other than at the corner hot spots
the electric fields remain finite, and thus there is no dissipation other
than at these corners.

The voltage drop in regime $(ii)$ can be calculated using a simple network
model, illustrated in Fig.~(4). (The hot spots where dissipation takes place
are marked $A$ in this figure.) The boundaries between different Hall
regions, or between a Hall region and the vacuum, are represented by bonds
in the model. Each bond is assigned a directional arrow, oriented so that
the Hall conductance $\sigma _{xy}=1/\rho _{yx}$ of the region on the
left-hand side of the bond is algebraically larger than the Hall conductance
on the right. (An insulating region is assigned a Hall conductance $\sigma
_{xy}=0$.) For each bond $\mu $, we denote the absolute value of the
difference in these Hall conductivities by $\sigma _{\mu }>0$. Note that if
one were to reverse the sign of the magnetic field, the signs of the Hall
conductances would change, as would the directions of the arrows, but $%
\sigma _{\mu }$ would be unchanged.

\begin{figure}
\includegraphics[width=3.375in,angle=0,clip=]{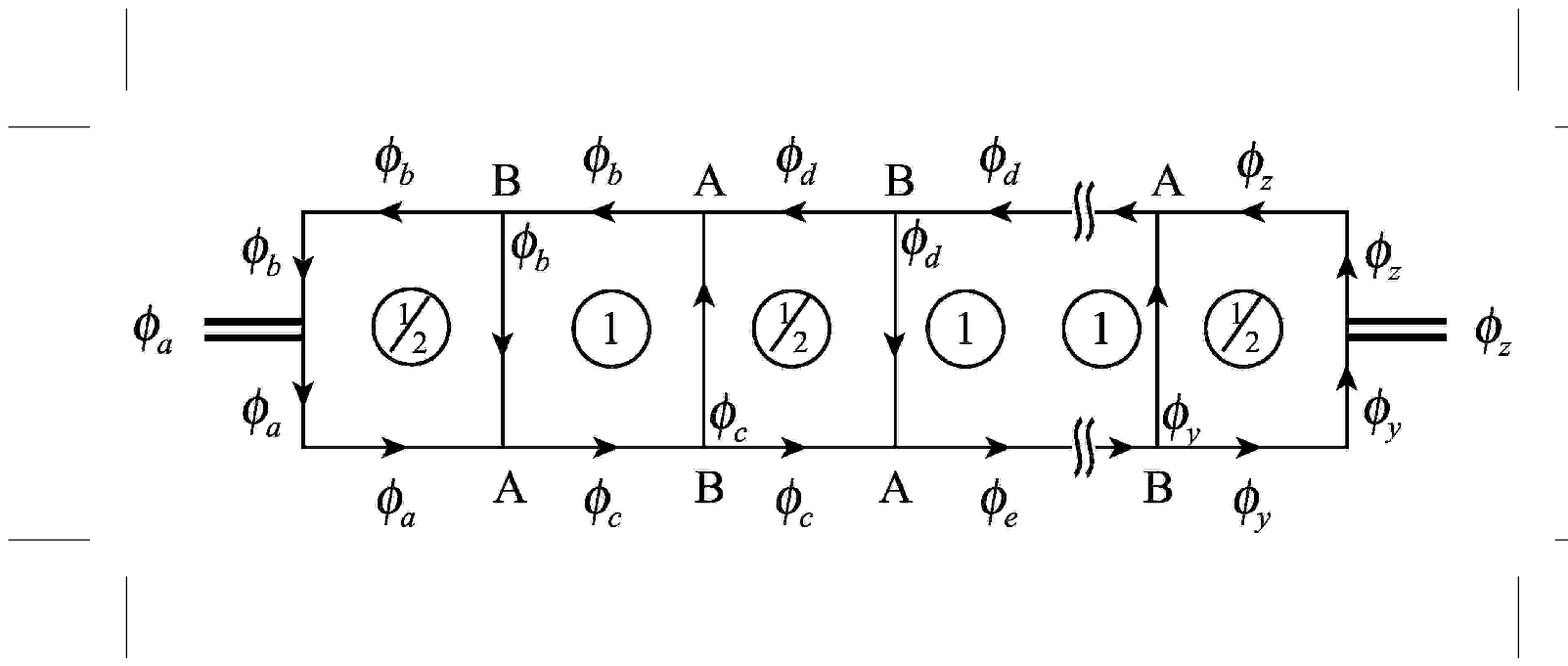}
\caption{\label{4}Network model when there is no dissipation in the interior of the
sample. Lines with arrows are bonds, with orientation as described in the
text. Double lines are metallic leads. Numbers in circles denote the Hall
conductivity $\sigma_{xy}$ (in units of $e^2/h$) within each region.
Potentials on the bonds, in the primary layer 1, and on the leads, are
denoted by $\phi_a, \phi_b,$ etc. Dissipation occurs at nodes labeled $A$
and at the contacts to the leads.}
\end{figure}

Let $\phi_{\mu} $ denote the potential $\phi^1$ on bond $\mu$, which is a
constant along the length of the bond. In the network model, the bond
carries a current, $i_{\mu} = \sigma_{\mu} \phi_{\mu} $, and an energy flux $%
i_{\mu} \phi_{\mu}$, with positive signs denoting transport in the direction
of the arrow. Although the current flow in the original problem is not
actually confined to the edges and boundaries, the details of the flow are
irrelevant to a computation of the voltage drop along the edges. The voltage
difference between any two points on the boundaries of a given Hall region
is completely determined by the total Hall current crossing a line joining
the two points, if the diagonal resistivity is zero. Thus there is no error
introduced by associating the currents with the boundaries.

Nodes in the network, where three bonds come together, represent the meeting
point of three Hall regions, Current conservation requires that the current
leaving a node must equal the current entering, while energy conservation
dictates that the energy flux leaving the node be equal to or smaller than
the energy entering. There are two types of nodes. When there are two bonds
with arrows pointing into the node (labeled $\mu=1,2 $), and one pointing
out (labeled $\mu = 3$), then the potentials on the incoming bonds are
arbitrary, and the potential on the outgoing bond is determined by current
conservation:
\begin{equation}
\phi_3 = (\phi_1 \sigma_1 + \phi_2 \sigma_2) / \sigma_3,
\end{equation}
with $\sigma_3 = \sigma_1 + \sigma_2 $. On the other hand, for a node with
one incoming bond and two outgoing bonds, the requirements of current and
energy conservation dictate that the potentials on the outgoing bonds be
equal to the potential on the incoming bond.

In addition to the nodes described above, we include additional junctions to
represent an ohmic contact with a metallic lead. For an ideal contact, the
condition is that the potential on the bond leaving the contact is the same
as the voltage in the metallic lead. The potential on the incoming bond is
arbitrary. (For a non-ideal contact, one may include a series resistance
which leads to an additional voltage drop if there is net current flowing
into or out of the lead.)

Using the above rules, we see that the there are no potential differences
among the three bonds connected to each of the nodes labeled $B$ in
Fig.~[4]. However, there are voltage drops at the nodes marked $A$. If the
net current in the $x-$direction is $I_{1}$, then we must have $I_{1}=\nu
_{1}(\phi _{a}-\phi _{b})=(\phi _{c}-\phi _{b})=\nu _{1}(\phi _{c}-\phi
_{d}) $, etc. Then we find $\phi _{a}-\phi _{c}=\phi _{b}-\phi
_{d}=I_{1}R^{\ast }$, where
\begin{equation}
R^{\ast }=(\nu _{1}^{-1}-1).  \label{Rstar2}
\end{equation}

\subsection{Cross-over between regimes (i) and (ii)\label{xoversec}}

Comparing Eqs. (\ref{Rstar}) and (\ref{Rstar2}) , we see that the two
expressions for $R^{\ast} $ become equal when the aspect ratio $L_d / w$ is
of order $\epsilon$, suggesting that this is indeed the boundary between
regimes (i) and (ii), as claimed. In fact, one can obtain an exact
expression for $R^{\ast}$ that is valid throughout this crossover regime. To
do this, let us consider the problem where both $\epsilon$ and the aspect
ratio $L_d / w$ are very small, but with arbitrary ratio between them. It is
convenient to think of $L_d$ as fixed, with $w$ very large and $\epsilon$
very small. As discussed above for the case of regime (ii), the reduced
potential $V$ on the lower edge $(y=0)$ will be a constant, equal to $V_0$,
except for a small interval, of order $\epsilon L/d$ near the right corner,
at $x= L_d/2$, while on the upper edge $(y=w)$ we have $V=0$ except for a
small interval near the left corner, at $x= -L_d/2$.

We may solve Eqs. (\ref{lapv},\ref{bcleft2},\ref{bcright2}) by writing
\begin{eqnarray}
V(x,y)&=& V_0 - \frac{ x V_{0}}{L_{d}}\\
&+&\sum_{n}\left( A_{n}e^{-k_{n}y}
+B_{n}e^{k_{n}y}\right)\nonumber \sin k_{n}\left(
x+L_{d}/2\right) ,\nonumber \label{expansion}
\end{eqnarray}
where $n$ is summed over positive integers and $k_{n}=\frac{\pi n}{L_{d}}.$
The coefficients $A_{n},B_{n}$ are determined by the values of $V$ at $y=0$
and $y=L$. We see that as long as $n\ll \epsilon ^{-1}$, one has
\begin{eqnarray}
A_{n} &\approx &-\frac{V_{0}}{\pi n} (-1)^n  \nonumber \\
B_{n} &\approx &-\frac{V_{0}}{\pi n}e^{-k_{n}w}
\end{eqnarray}
For $n > \epsilon^{-1}$, the coefficients fall off more rapidly than $1/n$.

The total current flowing in the $x$-direction can be found most
conveniently by evaluating the integral (\ref{bccurrent}) at the center
line, $x=0$. The first term in (\ref{expansion}) gives rise to a ``bulk''
contribution , $I_{x}^{{\rm bulk}}=(V_{0}\epsilon w/L_{d})[(1-\nu
_{1}^{-1})^{2}+\epsilon ^{2}]^{-2}$ , which is the same as that obtained
earlier in region (i). The second term in (\ref{expansion}) gives rise to an
``edge current'', concentrated in regions of order $L_{d}$ near the upper
and lower edges, and falling off exponentially away from these edges. As we
are considering the situation where $w\gg L_{d}$, the edge current is
independent of $w$, and simply adds to the bulk current. In the limit $%
\epsilon \ll 1$, the edge current is derived entirely from the first term in
(\ref{jV}) ({\it i.e.}, the Hall term) and it leads to a total contribution $%
I_{x}^{{\rm edge}}=V_{0}/(\nu _{1}^{-1}-1)$, which is the same as the result
in regime (ii).\cite{fnbh} Thus we find, for $\epsilon \ll 1$ and $L_{d}\ll w
$, including the crossover region between regimes (i) and (ii), the
resistance $R^{\ast }$ of a depleted region is given by
\begin{equation}
\frac{1}{R^{\ast }}=\frac{1}{\nu _{1}^{-1}-1}+\frac{\epsilon w}{L_{d}((1-\nu
_{1}^{-1})^{2}+\epsilon ^{2})}  \label{xover}
\end{equation}

Although the edge current near the center line $x=0$ is spread out over a
region of height $\approx L_d$, we note that very close to the interfaces,
at $x=\pm L_{d}/2$, the current is concentrated in a smaller interval, of
height $\approx \epsilon L_d$, at the hot-spot corner.

\subsection{Regime (iii) $\protect\epsilon^{-1} \ll L_d/w $}

It is clear that the length-independent expression for $R^{\ast}$ given by (%
\ref{Rstar2}) must break down, if $L_d$ is sufficiently large. When $%
L_{d}/w>\epsilon ^{-1}$(regime $(iii)$), the resistance of each stage is
proportional to the longitudinal resistivity of the depleted region, and is
given by 
\begin{equation}
R^{\ast }=\epsilon L_{d}/w.  \label{rstar3}
\end{equation}
This regime is of no interest, if we want to create a device with small $%
R^{\ast }$.

The crossover between regimes (ii) and (iii) may also be solved analytically
by considering a depleted region where both $\epsilon$ and $w/L_d$ are very
small, but with arbitrary ratio between them. This problem may be solved by
a similar method to that used for the crossover between regimes (i) and
(ii). In fact the two problems may be related by a duality transformation,
in which the electric fields and the currents are interchanged, and the
spatial coordinates are rotated by 90 degrees. The final result now is that
the resistance $R^{\ast}$ is the sum of the results given by (\ref{Rstar2})
and (\ref{rstar3}) .

\subsection{Total resistance $R_N$}

Eqs. (\ref{Rstar},\ref{Rstar2},\ref{rstar3}) give $R^{\ast },$ the
resistance for each of the $(N-1)$ intermediate depleted regions, in the
three aspect-ratio regimes. To this we must add the resistance arising from
the ohmic contacts to the depleted end tabs in Fig.~2. If the lengths $%
L_{d}^{\prime }$ of the end tabs are such that the aspect ratio $%
L_{d}^{\prime }/w$ is in the intermediate regime, large compared to $%
\epsilon $ but small compared to $\epsilon ^{-1}$, then the resistance of
the end tabs may analyzed using the network model, as illustrated in
Fig.~[4]. We readily find that the combined added resistance of the two end
tabs is equal to $2\nu _{1}^{-1}-1$. This resistance is composed of $\nu
_{1}^{-1},$ the two-terminal resistance of the depleted system, and $\nu
_{1}^{-1}-1,$ the resistance associated with the interface between the
depleted and $(111)$ parts. Thus we find a total resistance in layer $1$ of
\begin{equation}
R_{N}=[(N-1)R^{\ast }+(2\nu _{1}^{-1}-1)]\frac{h}{e^{2}}\approx \lbrack
(N-1)R^{\ast }+3]\frac{h}{e^{2}}  \label{RNfinal}
\end{equation}
[Having arrived at the presentation of our final formulas, we now restore
the factor of $h/e^{2}.$] For a device containing two voltage tabs in regime
$(ii)$ and one $(111)$ region $(N=1$), this gives a resistance $R_{1}\approx
3h/e^{2}.$

We note then that in order to have $V_{2}>V_{1}$, with the
device in Fig.~[2], we must choose the aspect ratio $L_{d}/w$
of the intermediate depleted regions to be smaller than
$\epsilon $. For technical reasons, it may be
difficult to attach an ohmic contact to a very short end tab whose length $%
L_{d}^{\prime }$ is smaller than $\epsilon w$. On the other hand, it should
be possible to fabricate a device where the lengths $L_{d}$ of the
intermediate depleted regions are very short, by using narrow wires as top
gates. Although the additional resistance $R^{\ast }$ for each intermediate
stage is small compared to $h/e^{2}$ in this case, the total resistance
cannot be smaller than the value $\approx 3h/e^{2}$ for a single stage
device. From (\ref{RNfinal}), we see that in order to have a larger voltage
in the secondary than in the primary, even for very small values of $%
L_{d}/\epsilon w$, the number of stages $N$ must be $\geq 4$.

\subsection{Effect of finite current in secondary}

We now consider what happens if there is a finite current $I_{2}$ drawn from
the secondary. The linearity of the circuit means that we can write
\begin{eqnarray}
V_{1} &=&AI_{1}+CI_{2}  \label{V1I2} \\
V_{2} &=&BI_{1}-DI_{2}  \label{V2I2}
\end{eqnarray}
where $A=R_{N}$ and $B=Nh/e^{2}$ as given above, and $C$ and $D$ are
constants to be determined now. If we set $I_{1}=0$ with $I_{2}\neq 0$, we
have effectively interchanged the roles of the primary and secondary layers.
We see from this that $C=Nh/e^{2}$. Also, if the tabs for the contacts to
layer 2 have an aspect ratio between $\epsilon $ and $\epsilon ^{-1}$, we
see that $D$ will be $N$ times the resistance of the primary layer of a
single stage device: $D\approx 3Nh/e^{2}$. The output impedance of the
secondary is appropriately defined as
\begin{equation}
Z=-\left( \frac{\partial V_{2}}{\partial I_{2}}\right) _{I_{1}}=D.
\end{equation}
If the secondary circuit is closed by a load resistance $R$, we find
\begin{equation}
\frac{V_{2}}{V_{1}}=\frac{BR}{AR+AD+BC}.  \label{V2V1}
\end{equation}

\subsection{Structure in Fig.~[1]}

Finally we consider the structure shown in Fig.~[1], which we
may analyze in a manner similar to the above. If the aspect
ratios of the depleted regions are all in the intermediate
regime $\epsilon <L_{d}/w<\epsilon ^{-1}$, and we define
$I_{1}$ as the current in {\it each} primary strip, then we
obtain
the following results for the constants in Eqs.~(\ref{V1I2}-\ref{V2V1}): $%
A\approx 3h/e^{2},C=h/e^{2},B=Nh/e^{2},D\approx
(N+2)h/e^{2}$. In this case we obtain a voltage ratio
$V_{2}/V_{1}\approx N/3$, which exceeds unity provided $N\geq
4$, For large $N$, the output impedance is lower than that of
the geometry in Fig.~[2].

\section{Additional Remarks}

Throughout our analysis, we have assumed that the dimensions of the system
are large enough for us to use macroscopic constitutive relations for the
current and voltage in each region. For the geometry of Fig.~2, the most
critical requirement for our analysis is that the length $L_d$ of the
depleted region must be larger than the mean free path of quasiparticles in
the single layer phase, so that a macroscopic resistivity may be used. In
general, for a large but finite system, there will be corrections to the
macroscopic equations arising from the boundaries between regions which
could either increase or decrease the total resistance. However, the
boundary contribution to the resistance should decrease as the reciprocal of
the length of the boundary, so that the boundary contribution becomes
negligible in the macroscopic limit.

\section{Conclusions}

In summary, in this paper we have considered two possible geometries for a
quantum Hall bilayer device which should be able to act as a $dc$
transformer with voltage gain, and analyzed the current flow patterns in
these geometries. In particular, our analysis permits us to calculate the
voltage differences between any two points on the edge of the sample, in
either layer, given the total current flow in each layer. The methods we
have used for this analysis are applicable more generally, to composite
systems where all components of the systems are characterized by a
longitudinal resistivity much smaller than the Hall resistivity. Finally,
experiments to test our proposals for a $dc$ voltage step-up transformer,
and to measure the voltage drops in various geometries, would help
strengthen our understanding of the novel interlayer correlations and phase
coherence found in strongly coupled $\nu =1 $ bi-layer systems.

\section*{Acknowledgments}

The authors are grateful to James Eisenstein for several helpful
discussions, including comments on the manuscript. This work was supported
by NSF DMR-0196503 (SMG), DMR-0233773 (BIH), the US-Israel Binational
Science Foundation (AS and BIH), and the Israel Science Foundation (AS).


\begin{thebibliography}{20}
\bibitem{curlfootnote}  Because vortices carry charge, there can be
frozen-in vortices which give a curl to the supercurrent but these fixed
circulating currents can be ignored as they do not modify the total
transport current.
\bibitem{fnbh}  Taking into account the second term in Eq. (\ref{jV}), we
find that the leading correction to the edge current $I_x^{{\rm edge}}$ is a
factor $[1 + \frac{\epsilon \ln 2}{(1-\nu^{-1}) \pi}]$, in the limit $%
\epsilon \ll 1$.
\end{thebibliography}
\end{document}